\documentclass{article}
\usepackage{spconf,amsmath,graphicx}
\usepackage{amssymb}
\usepackage{hyperref}
\usepackage{multirow}
\usepackage{booktabs}       
\usepackage[table, svgnames, dvipsnames]{xcolor}
\usepackage[normalem]{ulem}
\definecolor{lightergray}{gray}{0.85}
\newcommand{\DHnote}[1]{{\color{red}{\bf DH: }#1}}
\newcommand{\DHedit}[1]{{\color{black}#1}}
\newcommand{\so}[1]{} 
\definecolor{mygreen}{HTML}{008000}
\newcommand{\JPnote}[1]{{\color{mygreen}{\bf JP: }#1}}
\newcommand{\JPedit}[1]{{\color{black}#1}}

\title{Fast-Slow Transformer for Visually Grounding Speech}
\name{Puyuan Peng, David Harwath}
\address{Department of Computer Science, The University of Texas at Austin \\
Austin, Texas, 78712, USA\\
\texttt{\href{mailto:pyp@utexas.edu}{pyp},\href{mailto:harwath@utexas.edu}{harwath}@utexas.edu}}
\begin{document}
%
\maketitle
\begin{abstract}
We present \textbf{Fa}st-\textbf{S}low \textbf{T}ransformer for \textbf{V}isually \textbf{G}rounding \textbf{S}peech, or FaST-VGS\footnote{code and model weights are available at \href{https://github.com/jasonppy/FaST-VGS-Family}{https://github.com/jasonppy/FaST-VGS-Family}}. FaST-VGS is a Transformer-based model for learning the associations between raw speech waveforms and visual images. The model unifies dual-encoder and cross-attention architectures into a single model, reaping the superior retrieval speed of the former along with the accuracy of the latter. FaST-VGS achieves state-of-the-art speech-image retrieval accuracy on benchmark datasets, and its learned representations exhibit strong performance on the ZeroSpeech 2021 phonetic and semantic tasks.
\end{abstract}
\vspace{-2mm}
\section{Introduction and Related Work}\label{sec:intro}
\vspace{-2mm}
Humans acquire \so{knowledge}\DHedit{spoken language long before they can read and write. They do so} in a multimodal fashion, \so{among which hearing and seeing are the two most crucial ways.}\DHedit{by learning to relate the speech patterns they hear with sensory percepts occurring in other modalities, such as vision.\so{~\cite{Chen Yu's papers, others?}.\JPnote{I'm not familiar with Chen Yu's work, do you have recommended papers to cite?}}} Recently, a number of works on \so{speech-image} learning \DHedit{from visually-grounded speech} \so{has}\DHedit{have} \so{been proposed addressing}\DHedit{addressed} the question of whether machines can acquire knowledge \DHedit{of spoken language in a similar fashion~\cite{chru2021visually}}. \so{by hearing and seeing like humans do.} Researchers have investigated the benefits of incorporating visual context into automatic speech recognition~\cite{Sun16LookListenDecode,Palaskar18GroundedASR}, as a pre-training task for supervised ASR~\cite{Hsu2019TransferLF}, word detection and localization~\cite{Kamper2017VisuallyGL,Merkx2019LanguageLU,Wang2020ADH,Olaleye2021AttentionBasedKL}, hierarchical linguistic unit analysis~\cite{Chrupaa2017RepresentationsOL,Harwath2020LearningHD}, cross-modality alignment~\cite{Harwath2019JointlyDV,Wang2021AlignOA,khorrami2021evaluation}, speech segmentation~\cite{Harwath2019TowardsVG}, speech generation~\cite{Hsu2020TextFreeIS}, multilingual spoken language learning~\cite{Harwath2018VisionAA,Kamper2018VisuallyGC,Havard2020CatplayinginthesnowIO,Ohishi2020TrilingualSE}.

\DHedit{Many of these works have used retrieval between images and their spoken audio captions as a training task to learn speech representations from visual supervision. Although retrieval accuracy was often used as an evaluation benchmark to assess how well a model can predict visual semantics directly from a raw speech signal, in many cases these papers put a greater emphasis on analyzing how linguistic structure emerged within the representations learned by the model. In general, the accuracy of speech-image retrieval systems has lagged behind their text-image counterparts, but recently several works have made enormous progress towards closing this gap, demonstrating that speech-enabled image retrieval is a compelling application in its own right}\so{That said, most of the time this retrieval task was On the other hand, the task of speech-image retrieval, originally being used as a proxy task for examining the speech-image learning models, has been studied in isolation recently Recently, several works have }~\cite{Chrupaa2019SymbolicIB,Mortazavi2020SpeechImageSA,Ilharco2019LargescaleRL,Sanabria2021TalkDW}.\so{People  have shown that using massive external paired speech-image training data, direct speech-image retrieval models can outperform text-image retrieval models where text is provided by an ASR system~\cite{Sanabria2021TalkDW}.}

In this work, we propose \textbf{Fa}st-\textbf{S}low \textbf{T}ransformer for \textbf{V}isually \textbf{G}rounding \textbf{S}peech, or FaST-VGS, \DHedit{the first Transformer~\cite{Vaswani2017AttentionIA}-based model for speech-image retrieval. FaST-VGS uses a }\so{along with a} novel \so{end-to-end }coarse-to-fine training and retrieval method, which \DHedit{unifies the dual-encoder and cross-attention architectures,} \so{enables} \DHedit{enabling both} fast and accurate retrieval using a \so{unified}\DHedit{single} model. FaST-VGS achieves state-of-the-art speech-image retrieval accuracy \so{in}\DHedit{on the} Places Audio~\cite{Harwath2016UnsupervisedLO}, the Flickr8k \DHedit{Audio Caption Corpus (FACC)}~\cite{Harwath2015DeepMS}, and SpokenCOCO~\cite{Hsu2020TextFreeIS} benchmark corpora. In addition, we study the \so{phonetic and semantic properties of } \DHedit{linguistic information encoded in} the speech representations learned by FaST-VGS by \so{examining}\DHedit{evaluating} it on the phonetic and semantic tasks of the  ZeroSpeech 2021 Challenge~\cite{zs21} \DHedit{Visually-Grounded Track}~\cite{alishahi2021zr2021vg}.

\vspace{-3mm}
\section{Technical Approach}
\vspace{-2mm}
\begin{figure*}
  \centering
      \includegraphics[width=1\textwidth]{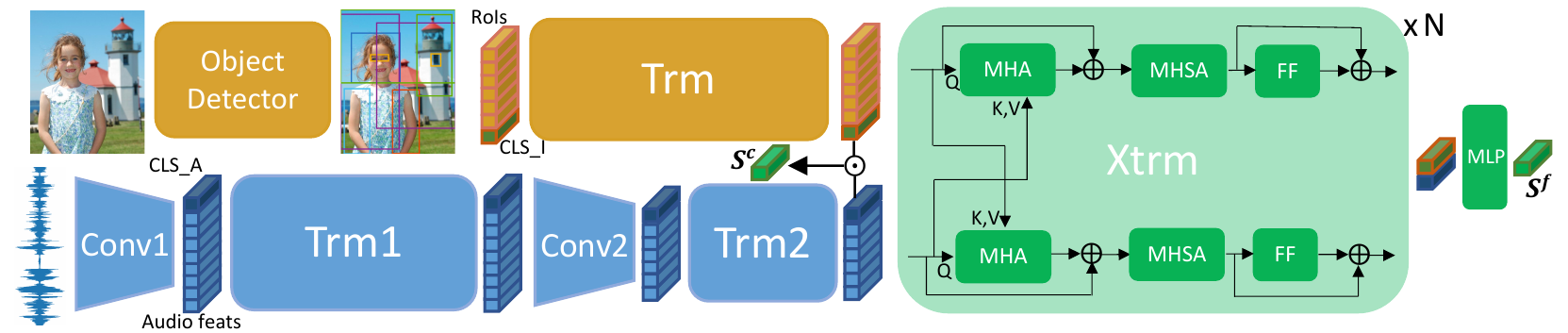} 
      \vspace{-.35in}
      \caption{Architecture of the FaST-VGS model.}\label{fig:archi}
      \vspace{-.2in}
\end{figure*}
\so{As shown in Fig.~\ref{fig:archi}, the }\DHedit{The FaST-VGS model is designed to take as input a visual image $I$ and an acoustic speech waveform $A$ that describes the contents of $I$, and output a similarity score that is large when $A$ faithfully describes the content of $I$, and small otherwise. Crucially, we design the model to output \textit{two} different similarity scores. The \textit{coarse} similarity $S^c$ is simply the dot product between independently embedded vector representations of $I$ and $A$, enabling computationally fast retrieval at the cost of reduced accuracy. Alternatively, the model can output a \textit{fine} similarity $S^f$ that leverages cross-modal self-attention to jointly encode $I$ and $A$, leading to more accurate but slower retrieval. Our} \DHedit{model (Fig.~\ref{fig:archi})} is composed of three modules: an audio encoder, an image encoder, and a cross-modal encoder\DHedit{, which we discuss in detail.}

The audio \so{branch}\DHedit{encoder takes as input the raw waveform of $A$, represented as a 1-D vector of sample amplitudes. It passes this waveform through a convolutional block (Conv1) to obtain features at a}\so{has four components, convent1 (Conv1), Transformer1 (Trm1), Conv2, and Trm2. Conv1 takes raw waveform as input and extracts} high \DHedit{temporal} resolution.\so{features.} Then, these \so{extracted audio} features \so{is} \DHedit{are augmented} \so{concatenated} with a \texttt{CLS\_A} token\so{along the temporal dimension. The result is} and input to \DHedit{a Transformer stack} (Trm1) for contextualization. The contextualized \DHedit{high resolution} \so{audio} features \so{output from Trm1} \DHedit{(with the exception of the \texttt{CLS\_A} token)} are \DHedit{then} fed into \DHedit{a second convolutional network} (Conv2) \so{for} \DHedit{in order to reduce their} temporal resolution. \DHedit{This is done to better capture higher level semantic information carried over a longer time span (e.g. by words and phrases), as well as to make the subsequent application of cross-modal attention less computationally expensive.} \so{reduction in order to it computationally tractable for later cross-modal processing.}The output of the Conv2 network is concatenated back with the \texttt{CLS\_A} token, and then fed as input to a second Transformer stack (Trm2) to perform another round of contextualization. At this point, the \texttt{CLS\_A} output of the Trm2 block can be directly used to compute the coarse similarity $S^c$, or all of the outputs can be carried onwards into the cross-modal Transformer block in order to compute the fine similarity $S^f$.\so{The bypassed \texttt{CLS\_A} will be concatenated with the output Conv2 and they together will be feed to Trm2 for further contextualization for more accurate coarse matching as will be introduced in detail later.} The image \so{branch}\DHedit{encoder} contains two components. First, an object detector\so{ (frozen during training)} extracts \DHedit{a set of} regions of interest (RoI) \DHedit{accompanied by their} features and \so{their} positions. Then, similar to the audio branch, a \texttt{CLS\_I} token is concatenated with RoI and position features which are input to a Transformer stack (Trm). \DHedit{At this point, the output \texttt{CLS\_I} token from the Trm block can be used with the \texttt{CLS\_A} token to compute the coarse score $S^c$, or fed with the contextualized RoI features into the cross-modal transformer to compute the fine score $S^f$.}
\so{The dot product of \texttt{CLS\_A} and \texttt{CLS\_I} output by the audio and image branch is taken to produce a similarity score which we refer to as the coarse similarity score, denoted by $S^c$, where $c$ stands for `coarse'. coarse similarity score can be calculated relatively quickly as no inter-modality interaction is required, but it can be inferior in terms of retrieval accuracy when being used as similarity measure, compared to the similarity calculated with inter-modal interaction that will be introduced later.} \DHedit{Finally, the cross-modal encoder (Xtrm) takes as input the outputs of the Trm and Trm2 blocks and passes them through a series of cross-modal Transformer blocks that perform contextualization over both input modalities at once.} \so{is also a Transformer which is the same as the one used in}\DHedit{For the Xtrm module, we use a slightly modified version of the Cross-Modality Encoder introduced by~\cite{Tan2019LXMERTLC}, in which each block is composed of a multi-head cross-attention mechanism, followed by a multi-head self attention mechanism and finally two feedforward layers. Residual connections are added around the cross-attention and feedforward layers.}\so{Each layer is composed of cross-attention, self-attention and fully connected layers. In Fig.~\ref{fig:archi} we show propagation mechanism of one cross-modal layer, and in cross-modal encoder such layer is repeated N times.} \DHedit{Differently from~\cite{Tan2019LXMERTLC}, our} \so{The} cross-modal encoder outputs only \DHedit{the} \texttt{CLS\_A} and \texttt{CLS\_I} \DHedit{tokens}, which are then concatenated and passed to a three-layer MLP to produce a scalar \DHedit{representing} \DHedit{$S^f$}. \so{(we also tried bilinear layer, and found that MLP works better).} \so{We refer to similarity score produced this way as the fine similarity score and denote it by $S^f$.}

\so{In the previous section, we have introduced the model architecture and similarity scores calculated using \texttt{CLS\_A} and \texttt{CLS\_I} from different part of the model for speech-image retrieval. We name retrieval using coarse matching score $S^c$ as coarse retrieval, and retrieval using fine matching score $S^f$ as fine retrieval.}\DHedit{To perform retrieval, we compute similarity scores between an input query (image or spoken description) and all of the items in the held-out target dataset, and return the $K$ items with the largest scores. Fine scores are more expensive to compute than the coarse scores, since the query must be paired with each target item and passed through the cross-modal attention module. In contrast, the embeddings of the target items used to compute the coarse scores may be pre-computed and re-used for each query. However, we expect the coarse scores to offer worse performance than the fine scores since they cannot leverage cross-modal attention.}\so{ \so{The have different speed-accuracy trade-off (we}\DHedit{We experimentally} confirm \so{this}\DHedit{these conjectures} \so{by experiments}in Sec.~\ref{sec:experiments}).} We can simultaneously achieve high speed and accuracy by using coarse-to-fine retrieval (CTF). \DHedit{CTF retrieval uses a two-pass strategy, where coarse retrieval is first used to retrieve a set of $K_c$ target items, where $K_c$ is set to be significantly larger than the $K$ items we ultimately want the model to return, but much smaller than the full size of the target dataset.\so{\so{The idea can be easily explained using an example, say the the query modality is audio, and we want the model to retrieve images that match the audio.} We pass images and audio to corresponding unimodal branches and get coarse similarity scores. We then sort the scores in descending order and select the top k images accordingly. }\so{These k}} \DHedit{The coarse feature outputs of the $K_c$ target items} \so{images' features} along with the \so{audio} \DHedit{coarse} features \DHedit{of the query} are then passed to \DHedit{the} cross-modal encoder to produce fine similarity scores, which \so{is}\DHedit{are} used for final ranking \DHedit{and the return of the top $K$ target items}.\so{ \DHedit{When pre-computation and caching of the coarse features is used, the} CTF retrieval method \so{turns}\DHedit{reduces} the time complexity of \DHedit{the fine-grained ranking step} \so{the computation in cross-modal encoder} from linear \so{in} (\DHedit{with respect to the size of the target}\so{search} database) \so{size }to \DHedit{a} constant. \so{in search database size, }\DHedit{While the coarse retrieval step will exhibit a linear time complexity with respect to the size of the target dataset (although this could be further reduced using approximate algorithms such as Locality Sensitive Hashing~\cite{lsh}), assuming the coarse features have been cached the the retrieval step consists of a single matrix multiplication followed by a sorting operation, and is thus much faster than performing a separate forward pass through the model for each target item.}} Our experiments in Sec.~\ref{sec:experiments} \so{and we will} confirm \so{by experiments in Sec.~\ref{sec:experiments}} that CTF can \so{significantly} reduce retrieval time \DHedit{by} \so{(}as much as 98\%\so{ time reduction)} compared to fine retrieval, while maintaining \DHedit{nearly the same} \so{it's}accuracy. ~\cite{Miech2021ThinkingFA} recently proposed a similar idea for text-to-image retrieval, where \so{they trained}\DHedit{a cross-modal transformer teacher model was used to train} a dual encoder \DHedit{student} model. \so{and a cross-modal transformer model separately with the later provides extra teacher supervision to the former, and use the former for coarse retrieval and latter for fine retrieval.} Our work differs from theirs in that we use a unified model for both coarse and fine retrieval, \so{and as will be revealed in the next section, the our model}\DHedit{which} is trained end-to-end with a simple objective.

\so{To encourage the model to produce a high matching score for matched pairs and low matching score for unmatched pairs, }For training, we follow~\cite{Ilharco2019LargescaleRL} and use the masked margin softmax loss that is simultaneously applied in both possible directions of retrieval within a batch. \so{Let}\DHedit{Given a} batch \so{size to be}\DHedit{of} $B$ \DHedit{pairs of images and audio captions we first compute the similarity score $S^{*}_{i,j}$ between the $i^{th}$ audio caption and the $j^{th}$ image, where $S^*$ can be either the coarse matching score $S^c$ or the fine matching score $S^f$.} \so{the}\DHedit{For a given matching score type, the} \so{matching} loss is \DHedit{then defined as:} $\mathcal{L}^* = \mathcal{L}^*_{A\rightarrow{I}} + \mathcal{L}^*_{I\rightarrow{A}},$ where  
\vspace{-2mm}
\begin{eqnarray}
    \mathcal{L}^*_{A\rightarrow{I}} = -\frac{1}{B}\sum_{i=1}^{B}\log\frac{e^{S^*_{i,i}-\delta}}{e^{S^*_{i,i}-\delta} + \sum_{j=1}^{B}M_{i,j}e^{S^*_{i,j}}} \\
    \mathcal{L}^*_{I\rightarrow{A}} = -\frac{1}{B}\sum_{i=1}^{B}\log\frac{e^{S^*_{i,i}-\delta}}{e^{S^*_{i,i}-\delta} + \sum_{j=1}^{B}M_{j,i}e^{S^*_{j,i}}}
\end{eqnarray}
\so{To ease the notation, we follow the convention that for all subscripts with $i, j$, the first index is always for audio and the second is always for images. }We assume \DHedit{that when $i = j$ the} audio \DHedit{caption} and image \so{with the same index} are matched, but both FACC and SpokenCOCO \DHedit{contain} multiple audio \DHedit{captions for each image.}\so{that match it.} \DHedit{In this case, it is possible that within a batch, certain images will have a ground-truth match with multiple audio captions. To prevent these from being treated as negative examples by the loss function,} \DHedit{ we use masking variables $M_{i,j}$ which are $0$ when audio caption $i$ matches image $j$, and 1 otherwise.} \so{$0-1$ variable which equals to $1$ if and only if audio $j$ match image $i$. $*$ is a placeholder for $c$ or $f$, which stand for `coarse' or `fine' respectively. With this notation, $S^c_{i,j}$ is the coarse similarity score between audio $i$ and image $j$.
$S^f_{j,i}$ is the fine similarity score between audio $j$ and image $i$. $M_{j,i}$ is a $0-1$ variable which equals to $1$ if and only if audio $j$ match image $i$.}\so{ $\delta$ is \so{the}\DHedit{a} margin \DHedit{hyperparameter} which \so{is chosen by us and}\DHedit{we set to 1 and leave} fixed during training.} We \so{training objective of FaST-VGS is the following:}\DHedit{train the model to optimize both the coarse and fine matching losses simultaneously using the overall loss} $\mathcal{L} = \lambda^c\mathcal{L}^{c} + \lambda^f\mathcal{L}^{f}$.
\vspace{-2mm}
\section{Experiments}
\label{sec:experiments}
\vspace{-2mm}
\textbf{Datasets}. We train and test our models on the Places Audio~\cite{Harwath2016UnsupervisedLO}, SpokenCOCO~\cite{Hsu2020TextFreeIS}, and Flickr8k Audio Captions~\cite{Harwath2015DeepMS} (FACC) datasets. Places Audio contains 400k images from the the MIT Places 205~\cite{Zhou2014LearningDF} dataset and each image has a \DHedit{single} corresponding audio caption \DHedit{that was recorded spontaneously by a human as they viewed the image}.\so{ that describes it.} Both SpokenCOCO and FACC were produced by having humans speak aloud the text captions accompanying the MSCOCO~\cite{Lin2014MicrosoftCC} and Flickr8k~\cite{Hodosh2013FramingID} datasets, respectively. SpokenCOCO contains 123k images, each with 5 spoken captions, while FACC contains 8k images each with 5 spoken captions. We use the standard train/dev/test splits for FACC, the Karpathy splits~\cite{Karpathy2017DeepVA} for SpokenCOCO, and the ``2020'' splits for Places Audio. \DHedit{Finally, we use the publicly-distributed development set of the ZeroSpeech 2021 Challenge~\cite{zs21, alishahi2021zr2021vg} to evaluate the phonetic and semantic information learned by our model.}

\begin{table}[htb]
\vspace{-3mm}
 \caption{\so{Ablation study of our model}\DHedit{Performance of} FaST-VGS on \DHedit{the} Places \DHedit{Audio}, averaged between the 2020 ``dev-seen'' and ``dev-unseen'' splits as well as between $A \rightarrow I$ and $I \rightarrow A$ retrieval.\so{ with sc which contains 1000 speech segments and 1000 corresponding images. R.M stands for retrieval methods, CTF stands for coarse-to-fine retrieval, and R.T. stands for retrieval time for one query averaged over speech and image on $8$ V100 GPUs. Retrieval accuracies are averaged over seen/unseen and image to speech and speech to image \DHnote{I think we can remove the trivial rows from this table, e.g. (0, 1) for coarse, (1, 0) for fine, and (1,0) and (0,1) for CTF} \JPnote{Agreed.}}}
 \vspace{-5mm}
\newcolumntype{g}{>{\columncolor{lightergray}}c}
\begin{center}
   \resizebox{\columnwidth}{!}{%
  \begin{tabular}{lcggggg}
      \toprule
      \rowcolor{white}
      Method & Time/Query &$\lambda^c$&$\lambda^f$&R@1&R@5&R@10\\
      \midrule
      \rowcolor{white}
      &&1& 0 &56.9 &85.2 &91.4 \\
      \rowcolor{white}
      & &1& 1 &60.7 &87.3 &93.0 \\
      \rowcolor{white}
      & &1& 0.1 &57.0 &85.3 &91.7 \\
      \multirow{-4}{*}{Coarse}&\multirow{-4}{*}{6.1ms}&0.1& 1 &61.8 &87.7 &93.3 \\
      \midrule
      \rowcolor{white}
      &&0& 1 &66.8 &89.1 &93.7\\
      \rowcolor{white}
      & &1& 1 &63.0 &88.3 &93.4 \\
      \rowcolor{white}
      &&1& 0.1 &56.5 &84.7 &91.7 \\
      \multirow{-4}{*}{Fine}&\multirow{-4}{*}{83.3ms}&0.1& 1 &67.8 &90.5 &94.7 \\
      \midrule
      \rowcolor{white}
      &&1& 1 &63.4 &88.3 &93.5 \\
      \rowcolor{white}
      & &1& 0.1 &56.7 &84.9 &91.5 \\
      \multirow{-3}{*}{CTF}&\multirow{-3}{*}{12.2ms}&0.1 & 1 &67.8 &90.5 &94.6 \\
      \bottomrule
  \end{tabular}}
  \end{center}
  \label{tab:abla_places}
  \vspace{-5mm}
  \end{table}
\textbf{Implementation details}. For the audio branch, we use the convolutional encoder and first 8 layers of the Transformer stack from wav2vec2.0~\cite{Baevski2020wav2vec2A} as the Conv1 and Trm1 blocks in our model. We initialize the weights of both blocks with the pretrained ``Base'' model, and we freeze the Conv1 weights during training. \so{Trm1 is a 8 layer Transformer, i.e. only the first $8$ layers of the transformer's weights of wav2vec 2.0 base is used and}We \so{find}\DHedit{found that using only the first 8 wav2vec2.0 transformer layers} \so{it} performs better than using all 12 layers.\so{ Conv2 and Trm2 are randomly initialized.} Conv2\so{has} \DHedit{uses the ResDAVEnet} architecture~\cite{Harwath2019JointlyDV}\so{ the same as~\cite{Hsu2019TransferLF}} with \so{only}the dimension of first \DHedit{convolutional} layer changed to \DHedit{match} Trm1's hidden dimension. Trm2 is a one layer standard Transformer, \DHedit{and both Conv2 and Trm2 are randomly initialized}. For \DHedit{the} image \so{branch}\DHedit{encoder}, we use \DHedit{the} off-the-shelf Faster-RCNN~\cite{Ren2015FasterRT} \so{distributed by~\cite{Tan2019LXMERTLC}}.\so{ by~\cite{Anderson2018BottomUpAT}.} \DHedit{We freeze Faster-RCNN weights, and} extract the top 36 RoIs for each image.\so{follow~\cite{Tan2019LXMERTLC} and use 36 features. Although the Faster RCNN is never trained on the three datasets we use here and recognition result is noisy, it still leads to good retrieval performance and speech representations.} The RoI features are fed to a 6-layer Transformer (Trm), which is randomly initialized. The cross-modal encoder \so{contains}\DHedit{consists of} \so{only }two \so{layers}\DHedit{blocks}, \DHedit{(N=2} in Figure~\ref{fig:archi}). All Transformer layers in our model use a hidden dimension of 768 and 8 attention heads \JPedit{(except for Trm1, which uses 12 attention heads)}.\so{, the cross-modal layer contains both self-attention and cross-attention (with hidden dimension and number of attention heads both the same as Trm2), a cross-modal layer contains two times more parameters than a unimodal Transformer layer.} \JPedit{The 3-layer MLP that computes $S^f$ has layers of width 1536, 768, and 1, and uses GELU~\cite{hendrycks2016gelu} activations.} To select $(\lambda^c, \lambda^f)$ we do a \so{three point}grid search over $\{0,0.1,1\}\times\{0,0.1,1\}$ on the Places Audio development set.\so{with redundant or meaningless combinations like \{0,0\} (meaningless), \{0.1,0.1\} (redundant), \{0,0.1\} (redundant) dropped.}\so{Interestingly, as will be shown in Sec.~\ref{sec:experiments}, $(\lambda^c, \lambda^f) = (0.1,1)$ achieves the highest retrieval accuracy on coarse retrieval, fine retrieval and CTF retrieval simultaneously. We provide discussion on the implications of this result in Sec.~\ref{sec:experiments}. }\so{ For optimization, we follow the standard practices for training Transformers, } All models are trained using BertAdam~\cite{Devlin2019BERTPO}\so{ (i.e. Adam with the bias-correction terms removed, and weight decay fixed\footnote{see  \href{https://github.com/huggingface/transformers/blob/694e2117f33d752ae89542e70b84533c52cb9142/README.md\#optimizers}{this description} on BertAdam}). We } with a linear warm-up \DHedit{over} the \so{learning rate for the }first $10\%$ \DHedit{of training} steps, a peak learning rate $1e-4$, and a linear decay to 0 \DHedit{over the course of training}. For all experiments, we set $\delta = 1$ and $K_c = 100$. Models on Places and SpokenCOCO are trained for $20$ epochs with batch size $48$ and $64$\so{ which are about $166$k and $177$k update steps respectively}, both taking less than $3$ days on $8$ NVIDIA V100 GPUs. Models on FACC are trained for $60$ epochs with batch size $96$\so{ which is about $19$k update steps}, and this takes about $0.5$ days on $8$ V100 GPUs.


\begin{table*}[htb]
\vspace{-3mm}
  \caption{Comparison of FaST-VGS with other models on the Places Audio Captions. Subscript CO means coarse retrieval.}
  \begin{center}
  \begin{tabular}{lcccccccccccc}
      \toprule
      &\multicolumn{6}{c}{Places Audio (test-seen)}&\multicolumn{6}{c}{Places Audio (test-unseen)}\\&\multicolumn{3}{c}{Speech $\rightarrow$ Image}&\multicolumn{3}{c}{Image $\rightarrow$ Speech}&\multicolumn{3}{c}{Speech $\rightarrow$ Image}&\multicolumn{3}{c}{Image $\rightarrow$ Speech}\\\cmidrule(lr){2-4} \cmidrule(lr){5-7} \cmidrule(lr){8-10} \cmidrule(lr){11-13} 
      Model&R@1&R@5&R@10 &R@1&R@5&R@10&R@1&R@5&R@10 &R@1&R@5&R@10\\
      \midrule
      ResDAVEnet~\cite{Hsu2019TransferLF} &35.2 &67.5& 78.0 &30.4 &63.1 &74.1 &38.3& 68.5& 78.8& 31.2& 65.0 &75.4\\
      MILAN~\cite{Sanabria2021TalkDW} & 58.4& 84.6& 90.6& 53.8 &83.4 &90.1 &62.1 &86.0 &90.5 &58.2 &85.8 &90.9 \\
      $\text{FaST-VGS}_{\text{CO}}$ & 60.0& 86.1& 92.3& 60.2 &85.1 &\textbf{92.2} &62.8 &88.4 &92.9 &62.3 &89.0 &93.2 \\
      $\text{FaST-VGS}_{\text{CTF}}$ & \textbf{64.0}& \textbf{88.3}& \textbf{93.7}& \textbf{64.2} &\textbf{87.9} &\textbf{92.2} &\textbf{69.6} &\textbf{90.3} &\textbf{94.3} &\textbf{66.0} &\textbf{90.4} &\textbf{94.1} \\
      \bottomrule
  \end{tabular}
  \end{center}
  \label{tab:place_main}
  \vspace{-8mm}
  \end{table*}

Table~\ref{tab:abla_places} displays the retrieval accuracy and time per query using coarse, fine, and CTF retrieval methods on the Places Audio dev set when varying the weights on the coarse and fine loss terms.\so{Tab.~\ref{tab:abla_places} shows how models trained with five different loss weights settings perform on Places Spoken Captions when using coarse retrieval, fine retrieval and coarse-to-fine retrieval. We first notice that f} \DHedit{F}or all three retrieval methods, $(\lambda^c,\lambda^f)=(0.1,1)$ achieves the highest\so{retrieving} scores. This \so{has two }indicat\DHedit{es}\so{ions,} \DHedit{that} \so{first, }even \DHedit{when}\so{if in the retrieval phase, only} coarse or fine \DHedit{matching are }\so{similarity score is }used \DHedit{alone}, \DHedit{training the model with both losses is}\so{using both coarse and fine matching loss can be} beneficial. \DHedit{Additionally,} \so{second,} putting more weight on \DHedit{the} fine matching loss is \DHedit{important}\so{crucial for the model to achieve better retrieval accuracy} regardless \DHedit{of the} retrieval method \DHedit{used}. \DHedit{In light of this, w}e fix $(\lambda^c,\lambda^f)=(0.1,1)$ for \DHedit{the remainder of our}\so{rest of} experiments. \DHedit{We also}\so{Another thing to} notice \so{is} that CTF maintains the accuracy of fine retrieval, while being \so{much faster}\DHedit{nearly as fast as coarse retrieval}. The \so{retrieval} speed gap is exacerbated when \DHedit{using a larger target}\so{the searching} database\DHedit{:}\so{ is larger,} on SpokenCOCO (which contains 5k images and 25k captions in the evaluation set)\so{ times more spoken captions than Places Audio Captions)}, \DHedit{fine and CTF retrieval achieve the same R@10 scores of $83.0$, but their average retrieval times per query are $399.7$ ms and $8.4$ ms.}\so{ the average retrieval time \DHedit{per query is}\so{for one sample anchor is} $399.7$ms using fine retrieval and $5.5$ ms using CTF retrieval, where CTF the retrieval accuracy is only decreased by 0.1\% absolute in R@10 when using CTF retrieval compare to fine retrieval.}
In Table.~\ref{tab:place_main}, we compare \DHedit{the retrieval} accuracy of FaST-VGS on the Places Audio test sets with other models \DHedit{from the literature} and show that both retrieval \DHedit{methods} significantly outperform MILAN~\cite{Sanabria2021TalkDW} which is the previous state-of-the-art \DHedit{for this task}.\so{ When using CO, the average retrieval accuracy at top 1 (R@1) is $61.3$ compare to $58.1$ for MILAN, and when using CTF, R@1 is $66.0$, which is $13\%$ higher than on MILAN.} We also note that the MILAN model was pretrained on 3.3 million image and spoken caption pairs from the Conceptual Spoken Captions (CSC)~\cite{Ilharco2019LargescaleRL} dataset, and hypothesize that training FaST-VGS on a large dataset would widen the performance gap further. 

\so{\so{Given the fact that in addition to Places training set, }MILAN \so{is also} Conceptual Spoken Captions (CSC)~\cite{Ilharco2019LargescaleRL}.\so{ which contains 3.3 million image and spoken caption pairs,} we believe the performance gap can be further increased if FaST-VGS is trained on same amount of data. We partially confirm this hypothesis in Tab.~\ref{tab:flickr8k_main}.}

In Table.~\ref{tab:flickr8k_main}, we \DHedit{present retrieval results on}\so{compare models on} FACC and SpokenCOCO. When trained only on FACC\so{(8000 images and 40000 spoken captions)}, FaST-VGS \DHedit{out}performs \DHedit{all previously published models trained on the same dataset, but still underperforms the }\so{worse than} MILAN \DHedit{model pretrained on CSC}.\so{, although still outperforms other models in the table. } To \DHedit{probe whether}\so{see if} this is \DHedit{due to the disparity in training data amounts,}\so{because of the dataset size,} we \DHedit{pretrain a}\so{load the weights of} FaST-VGS \DHedit{model}\so{ trained} on SpokenCOCO and \DHedit{fine-tune}\so{further train} it on FACC. \DHedit{This model significantly outperforms MILAN, despite the fact that the CSC dataset (3.3 million images and captions) is far larger than SpokenCOCO (123k images and 615k spoken captions).}\so{and results show that it outperforms MILAN. Note that even in this case the number of training data that FaST-VGS used (about 0.12 million images and 0.65 million spoken captions) is still much smaller than that of MILAN (3.3 million images and captions).} \DHedit{At the bottom of Table~\ref{tab:flickr8k_main} we present retrieval results on SpokenCOCO. The MILAN model preceded the SpokenCOCO dataset and hence did not report results on it, but we compare to the ResDAVEnet~\cite{Harwath2019JointlyDV} baseline model. We also compare to DIME~\cite{10.1145/3404835.3462829}, a current state-of-the-art model for image-text matching on MSCOCO, which can be considered an oracle upperbound for our model as DIME has access to the ground truth text being spoken. FaST-VGS significantly outperforms ResDAVEnet, and performs respectably against DIME in terms of average R@10 score (82.45 vs 86.25).} To test the performance of DIME on ASR text, we use the code and model weights open sourced by the authors and transcribe the SpokenCOCO test set using the Google ASR API. We denote this model as $\text{DIME}_{\text{ASR}}$ in Table.~\ref{tab:flickr8k_main}. The gap between FaST-VGS and $\text{DIME}_{\text{ASR}}$ is much smaller than $\text{DIME}$, and our $\text{FaST-VGS}_{\text{CTF}}$ performs equally or better in two of the six metrics compared to  $\text{DIME}_{\text{ASR}}$.
\begin{table}[htb]
\vspace{-3mm}
  \caption{Results on FACC and SpokenCOCO. $\dag$ indicates a model trained with text supervision. * indicates \so{that}\DHedit{a} \so{is}\DHedit{model} trained on SpokenCOCO and FACC\so{ Audio Captions}.}
  \vspace{-5.5mm}
  \begin{center}
  \resizebox{\columnwidth}{!}{%
  \begin{tabular}{lcccccc}
      \toprule
      &\multicolumn{3}{c}{Speech $\rightarrow$ Image}&\multicolumn{3}{c}{Image $\rightarrow$ Speech}\\\cmidrule(lr){2-4} \cmidrule(lr){5-7}  
      Model&R@1&R@5&R@10 &R@1&R@5&R@10\\
      \midrule
      &\multicolumn{6}{c}{Flickr8k (test 1k)}\\
      \cmidrule(lr){2-7}
      	~\cite{Chrupaa2017RepresentationsOL}&5.5&16.3&25.3&-&-&-\\
	   ~\cite{Havard2020CatplayinginthesnowIO}&9.6&-&-&-&-&-\\
	   ~\cite{Chrupaa2019SymbolicIB}$^\dag$&-&-&29.6&-&-&-\\
	   ~\cite{Merkx2019LanguageLU}&12.7&34.9&48.5&16.0&42.8&56.1\\
	  ~\cite{Higy2020TextualSF}$^\dag$ &21.8&49.9&63.1&-&-&-\\
      $\text{FaST-VGS}_{\text{CO}}$ &26.6&56.4&68.8&36.2&66.1&76.5 \\
      $\text{FaST-VGS}_{\text{CTF}}$&\textbf{29.3}&\textbf{58.6}&\textbf{71.0}&\textbf{37.9}&\textbf{68.5}&\textbf{79.9} \\
    \cmidrule(lr){1-7}
     ~\cite{Ilharco2019LargescaleRL}&13.9 &36.8 &49.5& 18.2 &43.5 &55.8 \\
      MILAN~\cite{Sanabria2021TalkDW} & 33.2 &62.7 &73.9 &49.6& 79.2& 87.5 \\
      $\text{FaST-VGS}^*_{\text{CO}}$ &41.2&70.4&81.5&53.6&81.1&89.5 \\
      $\text{FaST-VGS}^*_{\text{CTF}}$&45.5&73.8&83.7&59.8&84.1&90.7 \\
    \midrule
      &\multicolumn{6}{c}{SpokenCOCO (test 5k)}\\
      \cmidrule(lr){2-7}
      ResDAVEnet~\cite{Hsu2019TransferLF} &17.3 &41.9 &55.0 & 22.0&50.6 &65.2 \\
      $\text{FaST-VGS}_{\text{CO}}$ & 31.8& 62.5& 75.0& 42.5 &73.7 &84.9 \\
      $\text{FaST-VGS}_{\text{CTF}}$ & \textbf{35.9}& \textbf{66.3}& \textbf{77.9}& \textbf{48.8} &\textbf{78.2} &\textbf{87.0} \\
    \cmidrule(lr){1-7}
      &\multicolumn{3}{c}{Text $\rightarrow$ Image}&\multicolumn{3}{c}{Image $\rightarrow$ Text}\\
      \cmidrule(lr){2-4} \cmidrule(lr){5-7}
      DIME~\cite{10.1145/3404835.3462829} &40.2 &70.7 &81.4&56.1& 83.2 &91.1  \\
      $\text{DIME}_{\text{ASR}}$&37.6&66.3&76.9&54.8&82.4&90.4\\
      \bottomrule
  \end{tabular}}
  \end{center}
  \label{tab:flickr8k_main}
  \vspace{-6mm}
  \end{table}
   \begin{table}[htb]
   \vspace{-5mm}
  \caption{Performance on the ZeroSpeech 2021 tasks. Both VG models (bottom) are trained on SpokenCOCO, while the non-VG models (top) are trained on Librispeech~\cite{Panayotov2015LibrispeechAA}. \so{Result of FaST-VGS is Max pooling over features of Trm2.}}
  \vspace{-5.5mm}
  \begin{center}
   \resizebox{\columnwidth}{!}{
  \begin{tabular}{lcccccc}
      \toprule
&\multicolumn{2}{c}{Semantic $\uparrow$}&\multicolumn{4}{c}{ABX $\downarrow$}\\
\cmidrule(lr){2-3}\cmidrule(lr){4-7}
Model&Syn.&Lib.&W-C&W-O&A-C&A-O\\
      \midrule
     ~\cite{Niekerk21}&4.29&7.69 &0.03&0.05&0.04&0.08 \\
     ~\cite{Chorowski2021InformationRF}&5.90&10.20&0.03&0.04&0.04&0.07 \\
      \midrule
      VG Baseline (h.b.) &9.60&15.09&0.06&0.07&0.07&0.11 \\
      FaST-VGS &\textbf{15.80}&\textbf{23.55}&\textbf{0.05}&\textbf{0.06}&\textbf{0.05}&\textbf{0.09} \\
      \bottomrule
  \end{tabular}
  }
  \end{center}
  \label{tab:zs21_sem}
  \vspace{-6mm}
  \end{table}

Finally, we investigate the speech representations learned \so{in}\DHedit{by} FaST-VGS by \DHedit{evaluating it}\so{examining its performance} on the \DHedit{phonetic and semantic} tasks of the ZeroSpeech 2021 Challenge \DHedit{dev set}. \JPedit{We use representations from layer $6$ of Trm1 and layer $1$ of Trm2 (max pooled) for phonetic and semantic tasks respectively}, which we found to outperform the other layers in the model. \so{In particular, due to the scope and space limit of this paper, we focus on the the phonetic and semantic task. Compared to the other two tasks which involves language modeling, phonetic and semantic task require very little hyperparameter tuning and directly examine meaningful aspects of the representations. Since we didn't make a submission to the challenge, only devset result is available.} In Tab.~\ref{tab:zs21_sem}, we see that FaST-VGS achieves \so{the highest} very high semantic scores \so{among all models} for both sythesized speech and speech from Librispeech. For the phonetic \DHedit{ABX} task, FaST-VGS outperforms \DHedit{the} high budget VG baseline~\cite{alishahi2021zr2021vg} in all four ABX evaluations, but underperforms best non-VG models. 
\vspace{-3mm}
\section{Concluding Discussion}
\vspace{-2mm}
We presented FaST-VGS, a Transformer-based model for visually-grounded speech-image retrieval that achieves state-of-the-art results across 3 benchmark datasets, and makes use of a novel coarse-to-fine training and retrieval strategy that unifies the dual-encoder and cross-attention architectures into a single model. We analyzed the model's learned representations on two of the ZeroSpeech 2021 tasks, confirming that they are strongly predictive of phonetic identity and lexical semantics. In our future work, we plan to investigate other linguistic information learned by the model, as well as ways of using the model's self-attention heads for word and sub-word unit discovery.
\bibliographystyle{IEEE}
\small
\vspace{-3mm}
\bibliography{main}

\begin{thebibliography}{10}

\bibitem{chru2021visually}
G.~Chrupała,
\newblock ``Visually grounded models of spoken language: A survey of datasets,
  architectures and evaluation techniques,''
\newblock {\em arXiv}, 2021.

\bibitem{Sun16LookListenDecode}
F.~Sun, D.~Harwath, and J.~Glass,
\newblock ``Look, listen, and decode: Multimodal speech recognition with
  images,''
\newblock in {\em SLT}, 2016.

\bibitem{Palaskar18GroundedASR}
S.~Palaskar, R.~Sanabria, and F.~Metze,
\newblock ``End-to-end multimodal speech recognition,''
\newblock in {\em ICASSP}, 2018.

\bibitem{Hsu2019TransferLF}
W.-N. Hsu, D.~F. Harwath, and J.~Glass,
\newblock ``Transfer learning from audio-visual grounding to speech
  recognition,''
\newblock in {\em INTERSPEECH}, 2019.

\bibitem{Kamper2017VisuallyGL}
H.~Kamper, S.~Settle, G.~Shakhnarovich, and K.~Livescu,
\newblock ``Visually grounded learning of keyword prediction from untranscribed
  speech,''
\newblock in {\em INTERSPEECH}, 2017.

\bibitem{Merkx2019LanguageLU}
D.~Merkx, S.~Frank, and M.~Ernestus,
\newblock ``Language learning using speech to image retrieval,''
\newblock in {\em INTERSPEECH}, 2019.

\bibitem{Wang2020ADH}
L.~Wang and M.~Hasegawa-Johnson,
\newblock ``A dnn-hmm-dnn hybrid model for discovering word-like units from
  spoken captions and image regions,''
\newblock in {\em INTERSPEECH}, 2020.

\bibitem{Olaleye2021AttentionBasedKL}
K.~Olaleye and H.~Kamper,
\newblock ``Attention-based keyword localisation in speech using visual
  grounding,''
\newblock in {\em INTERSPEECH}, 2021.

\bibitem{Chrupaa2017RepresentationsOL}
G.~Chrupała, L.~Gelderloos, and A.~Alishahi,
\newblock ``Representations of language in a model of visually grounded speech
  signal,''
\newblock in {\em ACL}, 2017.

\bibitem{Harwath2020LearningHD}
D.~F. Harwath, W.-N. Hsu, and J.~R. Glass,
\newblock ``Learning hierarchical discrete linguistic units from
  visually-grounded speech,''
\newblock in {\em ICLR}, 2020.

\bibitem{Harwath2019JointlyDV}
D.~F. Harwath, A.~Recasens, D.~Sur{\'i}s, G.~Chuang, A.~Torralba, and J.~Glass,
\newblock ``Jointly discovering visual objects and spoken words from raw
  sensory input,''
\newblock {\em IJCV}, 2019.

\bibitem{Wang2021AlignOA}
L.~Wang, X.~Wang, M.~Hasegawa-Johnson, O.~Scharenborg, and N.~Dehak,
\newblock ``Align or attend? toward more efficient and accurate spoken word
  discovery using speech-to-image retrieval,''
\newblock in {\em ICASSP}, 2021.

\bibitem{khorrami2021evaluation}
K.~Khorrami and O.~Räsänen,
\newblock ``Evaluation of audio-visual alignments in visually grounded speech
  models,''
\newblock in {\em INTERSPEECH}, 2021.

\bibitem{Harwath2019TowardsVG}
D.~F. Harwath and J.~R. Glass,
\newblock ``Towards visually grounded sub-word speech unit discovery,''
\newblock in {\em ICASSP}, 2019.

\bibitem{Hsu2020TextFreeIS}
W.-N. Hsu, D.~F. Harwath, C.~Song, and J.~Glass,
\newblock ``Text-free image-to-speech synthesis using learned segmental
  units,''
\newblock in {\em ACL}, 2021.

\bibitem{Harwath2018VisionAA}
D.~F. Harwath, G.~Chuang, and J.~R. Glass,
\newblock ``Vision as an interlingua: Learning multilingual semantic embeddings
  of untranscribed speech,''
\newblock {\em ICASSP}, 2018.

\bibitem{Kamper2018VisuallyGC}
H.~Kamper and M.~Roth,
\newblock ``Visually grounded cross-lingual keyword spotting in speech,''
\newblock {\em SLTU}, 2018.

\bibitem{Havard2020CatplayinginthesnowIO}
W.~N. Havard, J.-P. Chevrot, and L.~Besacier,
\newblock ``Catplayinginthesnow: Impact of prior segmentation on a model of
  visually grounded speech,''
\newblock in {\em CoNLL}, 2020.

\bibitem{Ohishi2020TrilingualSE}
Y.~Ohishi, A.~Kimura, T.~Kawanishi, K.~Kashino, D.~F. Harwath, and J.~Glass,
\newblock ``Trilingual semantic embeddings of visually grounded speech with
  self-attention mechanisms,''
\newblock {\em ICASSP}, 2020.

\bibitem{Chrupaa2019SymbolicIB}
G.~Chrupała,
\newblock ``Symbolic inductive bias for visually grounded learning of spoken
  language,''
\newblock in {\em ACL}, 2019.

\bibitem{Mortazavi2020SpeechImageSA}
M.~Mortazavi,
\newblock ``Speech-image semantic alignment does not depend on any prior
  classification tasks,''
\newblock in {\em INTERSPEECH}, 2020.

\bibitem{Ilharco2019LargescaleRL}
G.~Ilharco, Y.~Zhang, and J.~Baldridge,
\newblock ``Large-scale representation learning from visually grounded
  untranscribed speech,''
\newblock in {\em CoNLL}, 2019.

\bibitem{Sanabria2021TalkDW}
R.~Sanabria, A.~Waters, and J.~Baldridge,
\newblock ``Talk, don't write: A study of direct speech-based image
  retrieval,''
\newblock in {\em INTERSPEECH}, 2021.

\bibitem{Vaswani2017AttentionIA}
A.~Vaswani, N.~M. Shazeer, N.~Parmar, J.~Uszkoreit, L.~Jones, A.~N. Gomez,
  L.~Kaiser, and I.~Polosukhin,
\newblock ``Attention is all you need,''
\newblock in {\em NeurIPS}, 2017.

\bibitem{Harwath2016UnsupervisedLO}
D.~F. Harwath, A.~Torralba, and J.~R. Glass,
\newblock ``Unsupervised learning of spoken language with visual context,''
\newblock in {\em NIPS}, 2016.

\bibitem{Harwath2015DeepMS}
D.~F. Harwath and J.~R. Glass,
\newblock ``Deep multimodal semantic embeddings for speech and images,''
\newblock {\em ASRU}, 2015.

\bibitem{zs21}
T.~A. Nguyen, M.~de~Seyssel, P.~Rozé, M.~Rivière, E.~Kharitonov, A.~Baevski,
  E.~Dunbar, and E.~Dupoux,
\newblock ``The zero resource speech benchmark 2021: Metrics and baselines for
  unsupervised spoken language modeling,''
\newblock {\em SAS @ NeurIPS}, 2020.

\bibitem{alishahi2021zr2021vg}
A.~Alishahi et~al.,
\newblock ``Zr-2021vg: Zero-resource speech challenge, visually-grounded
  language modelling track, 2021 edition,'' 2021.

\bibitem{Tan2019LXMERTLC}
H.~Tan and M.~Bansal,
\newblock ``Lxmert: Learning cross-modality encoder representations from
  transformers,''
\newblock in {\em EMNLP}, 2019.

\bibitem{Miech2021ThinkingFA}
A.~Miech, J.-B. Alayrac, I.~Laptev, J.~Sivic, and A.~Zisserman,
\newblock ``Thinking fast and slow: Efficient text-to-visual retrieval with
  transformers,''
\newblock in {\em CVPR}, 2021.

\bibitem{Zhou2014LearningDF}
B.~Zhou, {\`A}.~Lapedriza, J.~Xiao, A.~Torralba, and A.~Oliva,
\newblock ``Learning deep features for scene recognition using places
  database,''
\newblock in {\em NeurIPS}, 2014.

\bibitem{Lin2014MicrosoftCC}
T.-Y. Lin, M.~Maire, S.~J. Belongie, J.~Hays, P.~Perona, D.~Ramanan,
  P.~Doll{\'a}r, and C.~L. Zitnick,
\newblock ``Microsoft coco: Common objects in context,''
\newblock in {\em ECCV}, 2014.

\bibitem{Hodosh2013FramingID}
M.~Hodosh, P.~Young, and J.~Hockenmaier,
\newblock ``Framing image description as a ranking task: Data, models and
  evaluation metrics (extended abstract),''
\newblock {\em JAIR}, 2013.

\bibitem{Karpathy2017DeepVA}
A.~Karpathy and L.~Fei-Fei,
\newblock ``Deep visual-semantic alignments for generating image
  descriptions,''
\newblock {\em TPAMI}, 2017.

\bibitem{Baevski2020wav2vec2A}
A.~Baevski, H.~Zhou, A.~Mohamed, and M.~Auli,
\newblock ``wav2vec 2.0: A framework for self-supervised learning of speech
  representations,''
\newblock in {\em NeurIPS}, 2020.

\bibitem{Ren2015FasterRT}
S.~Ren, K.~He, R.~B. Girshick, and J.~Sun,
\newblock ``Faster r-cnn: Towards real-time object detection with region
  proposal networks,''
\newblock {\em TPAMI}, 2015.

\bibitem{hendrycks2016gelu}
D.~Hendrycks and K.~Gimpel,
\newblock ``Gaussian error linear units (gelus),''
\newblock {\em ArXiv}, 2016.

\bibitem{Devlin2019BERTPO}
J.~Devlin, M.-W. Chang, K.~Lee, and K.~Toutanova,
\newblock ``Bert: Pre-training of deep bidirectional transformers for language
  understanding,''
\newblock in {\em NAACL}, 2019.

\bibitem{10.1145/3404835.3462829}
L.~Qu, M.~Liu, J.~Wu, Z.~Gao, and L.~Nie,
\newblock ``Dynamic modality interaction modeling for image-text retrieval,''
\newblock in {\em ACM SIGIR}, 2021.

\bibitem{Higy2020TextualSF}
B.~Higy, D.~Eliott, and G.~Chrupała,
\newblock ``Textual supervision for visually grounded spoken language
  understanding,''
\newblock in {\em Findings of EMNLP}, 2020.

\bibitem{Panayotov2015LibrispeechAA}
V.~Panayotov, G.~Chen, D.~Povey, and S.~Khudanpur,
\newblock ``Librispeech: An asr corpus based on public domain audio books,''
\newblock in {\em ICASSP}, 2015.

\bibitem{Niekerk21}
B.~v.~Niekerk, L.~Nortje, M.~Baas, and H.~Kamper,
\newblock ``Analyzing speaker information in self-supervised models to improve
  zero-resource speech processing,''
\newblock in {\em INTERSPEECH}, 2021.

\bibitem{Chorowski2021InformationRF}
J.~Chorowski et~al.,
\newblock ``Information retrieval for zerospeech 2021: The submission by
  university of wroclaw,''
\newblock {\em ArXiv}, 2021.

\end{thebibliography}

\end{document}